# Functional Superoscillations in a Wave Processor

Yuntian Wang[1,2,3], Md Sadman Sakib Rahman[1,2,3], Aydogan Ozcan[1,2,3,*]

[1]Electrical and Computer Engineering Department, University of California, Los Angeles, CA, 90095, USA [2]Bioengineering Department, University of California, Los Angeles, CA, 90095, USA [3]California NanoSystems Institute (CNSI), University of California, Los Angeles, CA, 90095, USA

*ozcan@ucla.edu

**Abstract**

Physical wave processors are constrained by diffraction, but the bandwidth of the mathematical functions they implement need not be set by it. Here, we demonstrate continuous and multiplexed optical processing of superoscillatory functions, in which a scalar argument is encoded onto a wavefront and a fixed diffractive wave processor synthesizes optical responses that approximate the corresponding function outputs in parallel. Unlike conventional superoscillatory optics that generates predetermined fields, superoscillation here is expressed in the input-output dependence on the encoded argument. Its frequencies are therefore distinct from the transverse spatial frequencies of propagating light and are not subject to the same diffraction-imposed cutoff. We demonstrate complex-valued superoscillatory function synthesis and introduce a multiplexed architecture that simultaneously evaluates distinct target superoscillatory functions whose harmonic content exceeds that of the optical input encoding. Visible-light experiments further validate this beyond-encoding superoscillatory regime. Together, these results establish functional superoscillation as a design principle for wave-based computation and show how interference, diffractive depth and spatial channel allocation extend the functional reach of passive wave processors. More broadly, these analyses establish a separation between functional bandwidth and physical wave bandwidth in analog optical computation.

**Introduction**

Superoscillation describes a phenomenon in which a band-limited function locally oscillates faster than its highest constituent Fourier component(*1–7*). Rather than introducing higher spatial or temporal frequencies, superoscillatory behavior arises from precisely structured interference among lower-frequency components. This property has attracted substantial interest in optics because it enables sub-diffraction focusing and imaging using only propagating waves, without relying on evanescent fields (*8–14*). Related concepts have also been explored in optical metrology(*15*), spectroscopy(*9*), particle manipulation(*16*, *17*), and signal processing(*18*). However, superoscillation carries a fundamental trade-off: faster local oscillations generally require increasingly large and finely balanced expansion coefficients, producing high-amplitude sidelobes and concentrating only a small fraction of the available energy within the superoscillatory region. These requirements create severe dynamic-range and sensitivity constraints, making strong superoscillatory functions difficult to synthesize accurately in physical systems(*17*, *19–21*).

Most optical implementations of superoscillations have focused on generating a predetermined spatial field, such as a sub-diffraction focal spot, through optimized masks, lenses, and metasurfaces(*22–32*). In these systems, a fixed optical element transforms an incident wave into a desired spatial intensity distribution, which is static. Superoscillatory function approximation, the topic of this work, presents a related but distinct problem. Instead of generating a single, static spatial pattern, we consider wave processors that receive a continuously varying query variable $t$ and return the corresponding value of a target superoscillatory function $F(t)$ while faithfully reproducing the rapid variations that define their superoscillatory behavior. This task of achieving functional superoscillations becomes particularly challenging when the target contains frequencies beyond the maximum frequency represented in the input encoding. Phase-encoded diffractive wave processors provide a platform for parallel function approximation through passive wave propagation and optical readout(*33*). However, functional superoscillations introduce a distinct, challenging question: how can a wave processor reproduce rapid local input–output variations without supplying corresponding high-frequency components in its native encoding? Addressing this question requires control over both the interference among the available harmonics and the allocation of optical degrees of freedom across the function domain. It connects the physics of superoscillations to the design of computational representation of a wave processor.

Here, we implement superoscillatory functions as dynamic optical input–output mappings: a continuously variable argument is encoded onto a wavefront, and a fixed wave processor returns the corresponding value of the target function. The superoscillatory frequencies in this problem are therefore frequencies with respect to the mathematical input argument, rather than spatial frequencies of propagating optical waves. Consequently, the wave processor remains fully governed by linear diffraction, while the function that it implements can exhibit nonlinear variations faster than the frequencies directly represented in its input encoding. Stated differently, this work establishes a distinct regime of superoscillations in which the superoscillation phenomenon resides in the functional dependence of the output on an abstract input argument, rather than in the spatial-frequency content of propagating waves.

We demonstrate this principle using multi-layer optical processors based on coherent wave propagation and intensity detection. First, we simultaneously approximate the real and imaginary parts of multiple complex-valued Berry–Popescu superoscillatory functions, whose constituent harmonics remain within the encoded frequency band while their local oscillation rates exceed this bandlimit, and show that increasing diffractive depth of the optical processor substantially improves the function approximation accuracy as the superoscillation strength increases. We then introduce a spatially multiplexed architecture that partitions the function domain among multiple output channels, enabling accurate evaluation of functions whose target harmonic content extends beyond that represented in the optical input encoding. In this architecture, each output channel is responsible for approximating the target over only a restricted sub-interval of the input domain, converting the global superoscillation encoding problem into a set of locally tractable mappings. We experimentally validate this approach using a visible-light set-up with a phase-only spatial light modulator (SLM) and *in situ* learning, recovering the target superoscillation function over a continuous input domain using an encoder containing fewer harmonics than the target. These results establish diffractive wave processing as a physical framework for continuous evaluation and synthesis of superoscillatory functions, distinct from the generation of static superoscillatory fields, and illustrate that functional bandwidth in a wave-based computing system need not be identified with the spatial-frequency bandwidth of the propagating waves that implement it.

## Results

We illustrate our diffractive superoscillatory function approximation framework in **Fig. 1**. As shown in **Fig. 1A**, we first construct a complex-valued target function as a weighted sum of $N_{\mathrm{sup}}$ harmonic basis functions, $f_j(t) = \exp(\mathrm{i}k_j t)$ with $j = 0,1,\ldots,N_{\mathrm{sup}} - 1$ and $\mathrm{i} = \sqrt{-1}$, giving $G(t) = \sum_{j=0}^{N_{\mathrm{sup}}-1} w_j f_j(t) = \sum_{j=0}^{N_{\mathrm{sup}}-1} w_j \exp(\mathrm{i}k_j t)$. The constituent wavenumbers $k_j$ are restricted to a finite interval, with $k_{\max} = \max_j |k_j|$, while the complex-valued weights $w_j$ are chosen such that, within the region of interest $[-t_{\mathrm{roi}}, t_{\mathrm{roi}}]$, $G(t)$ exhibits a local oscillation rate exceeding $k_{\max}$, characteristic of a superoscillatory function. For a nonzero complex-valued function $G(t)$, the local oscillation rate is defined as $k_{\mathrm{local}}(t) = \mathrm{Im}[G'(t)/G(t)]$ where $G'(t)$ refers to the first derivative of $G(t)$. The target superoscillatory function is obtained by restricting $G(t)$ to this region through multiplication with a rectangular window, $F(t) = G(t)[u(t + t_{\mathrm{roi}}) - u(t - t_{\mathrm{roi}})]$, where $u(\cdot)$ denotes the unit step function. Although the constituent harmonic content of $G(t)$ remains bounded by $k_{\max}$, their coherent superposition can vary locally at a substantially higher rate over a finite interval, which is the defining feature of superoscillations.

**Figure 1B** shows how this superoscillatory target function can be synthesized through a wave processor. We denote the number of harmonics available at the input encoder by $N_{\mathrm{base}}$ and define the target-to-encoding harmonic ratio as:

$$\rho_h = \frac{N_{\mathrm{sup}}}{N_{\mathrm{base}}} \tag{1}$$

In our analysis, we first consider $\rho_h = 1$, where the target function and input encoding share the same harmonic basis, while the case of $\rho_h > 1$ is addressed later through piecewise optical synthesis of superoscillations beyond the encoding basis, detailed in the next sub-section. The scalar input $t$ is mapped onto the optical encoder phase profile $\phi_{\mathrm{enc}}(t)$, in which each of the $\sqrt{N_{\mathrm{base}}} \times \sqrt{N_{\mathrm{base}}}$ input pixels carries a phase $\phi_j(t) = k_j t$. Under uniform coherent illumination, the corresponding input field can be expressed as:

$$U_{\text{in}}(\boldsymbol{r}_i;t) = \sum_{j=0}^{N_{\text{base}}-1} p_j(\boldsymbol{r}_i)\, e^{\mathrm{i}k_j t} \tag{2}$$

where $\boldsymbol{r}_i$ refers to the input plane coordinates, and $p_j(\boldsymbol{r}_i)$ is a unit-valued aperture function over the spatial extent of the $j$-th encoder pixel at the input plane and zero elsewhere, and the complete input field $U_{\text{in}}(\boldsymbol{r}_i;t)$ therefore contains the same harmonic dependence $\exp(\mathrm{i}k_j t)$ used to construct $G(t)$. The wave processor can be described by a spatially varying complex impulse response $h(\boldsymbol{r}_o,\boldsymbol{r}_i)$, relating an input coordinate $\boldsymbol{r}_i$ to an output coordinate $\boldsymbol{r}_o$. The output field can therefore be expressed as:

$$U_{\text{out}}(\boldsymbol{r}_o;t) = \int_{\mathcal{D}_i} h(\boldsymbol{r}_o,\boldsymbol{r}_i) U_{\text{in}}(\boldsymbol{r}_i;t) d\boldsymbol{r}_i$$

$$= \sum_{j=0}^{N_{\text{base}}-1} e^{\mathrm{i}k_j t} \int_{\mathcal{D}_i} h(\boldsymbol{r}_o,\boldsymbol{r}_i) p_j(\boldsymbol{r}_i) d\boldsymbol{r}_i = \sum_{j=0}^{N_{\text{base}}-1} H_j(\boldsymbol{r}_o) e^{\mathrm{i}k_j t} \tag{3}$$

Here, $H_j(\boldsymbol{r}_o)$ represents the complex-valued transfer function from the $j$-th encoded input region to the output coordinate $\boldsymbol{r}_o$ and $\mathcal{D}_i$ denotes the input plane of the wave processor. Therefore, a passive wave processor forms, at each output position, a distinct coherent superposition of the encoded harmonic components. The output field is subsequently converted into measurable intensity signals through square-law detection. For a detector occupying the region $\mathcal{D}_m$,

$$I_m(t) = \int_{\mathcal{D}_m} |U_{\text{out}}(\boldsymbol{r}_o;t)|^2 d\boldsymbol{r}_o = \sum_{p=0}^{N_{\text{base}}-1} \sum_{q=0}^{N_{\text{base}}-1} C_{p,q}^m e^{i(k_p-k_q)t} \tag{4}$$

where $C_{p,q}^m = \int_{\mathcal{D}_m} H_p(\boldsymbol{r}_o) H_q(\boldsymbol{r}_o)^* \, d\boldsymbol{r}_o$. For the complex-valued superoscillatory target functions, $F(t)$, considered here, two sets of differential detector pairs are used to recover the real (Re) and imaginary (Im) components through the wave processor (see **Fig. 1B**), giving:

$$\hat{F}(t) = [I_{\text{Re}}^+ - I_{\text{Re}}^-] + \mathrm{i}[I_{\text{Im}}^+ - I_{\text{Im}}^-] \tag{5}$$

where $t \in [-t_{\text{roi}}, t_{\text{roi}}]$. Thus, the superoscillatory function approximation problem can be viewed physically as finding a realizable complex impulse response whose spatially varying weights, $h(\boldsymbol{r}_o,\boldsymbol{r}_i)$, produce the required coherent interference among the encoded harmonics over the region of interest. We solve this inverse-design problem end-to-end by embedding the input encoding, free-space propagation, trainable phase-only diffractive layers, and intensity readout within a differentiable optical forward model(*34*). The diffractive phase profiles are optimized over batches of continuously sampled input arguments to minimize the discrepancy between the measured detector responses and the corresponding target superoscillatory function values, thereby learning the coherent complex-valued weights of the encoded harmonics required for superoscillatory function approximation, i.e., $\hat{F}(t) \approx F(t)$; see the **Methods** for details. By optimizing the detected response directly over the evaluation interval, this formulation makes local functional behavior the object of optical function synthesis. The learned interference pattern can therefore reproduce the required superoscillatory response without a coefficient-by-coefficient reproduction of the target functions' global Fourier expansion. In fact, as we detail later in the next sub-section, this framework can also be applied to $\rho_h > 1$ where target superoscillatory functions contain more harmonics than the native encoder (i.e., $N_{\text{sup}} > N_{\text{base}}$), achieving optical synthesis of superoscillations beyond the encoding basis.

We first demonstrate the approximation of Berry–Popescu (BP) superoscillatory functions(*14*) using a multi-layer diffractive wave processor (with $L = 4$ optimized phase-only layers). The system architecture and target functions are illustrated in **Fig. 2A**. For this BP demonstration, the number of target harmonics matches that of the input encoding, i.e., $N_{\text{sup}} = N_{\text{base}} = n + 1$, corresponding to $\rho_h = 1$. The target functions are denoted by $F_i\ (a_i, t)$, with $i = 1, \dots, N_f$, where $N_f$ is the number of distinct superoscillatory functions implemented in parallel and $a_i$ is the superoscillation parameter of function $i$ that determines the local oscillation frequency relative to the maximum constituent harmonic frequency. The underlying BP generating function is expressed as $G_i(a_i, n; t) = \sum_{j=0}^{n} A_{i,j} e^{\mathrm{i} k_j t}$ where $A_{i,j}$ denotes the analytical BP expansion coefficient associated with the constituent harmonic $k_j$; its closed-form expression is given in **Methods**. In this notation, $n + 1 = N_{\text{sup}} = N_{\text{base}}$ constituent wavenumbers satisfy $k_j \in [-1,1]$, and $G_i$ is therefore band-limited to $|k| \leq 1 = k_{\max}$. Within the input argument region of interest $t \in [-t_{\text{roi}}, t_{\text{roi}}]$, $G_i$ oscillates locally at the frequency $a_i$ near $t = 0$, exceeding its highest constituent harmonic by a superoscillation factor of $a_i/k_{\max} = a_i$. The target function $F_i$ is defined by restricting $G_i$ to this region, with $t_{\text{roi}} = 2$ unless otherwise mentioned.

In our initial design, the multi-layer wave processor is optimized to simultaneously implement $N_f = 100$ target complex-valued superoscillatory functions (parameterized by $a_i \in [1,5]$), recovering both the real and imaginary components via two sets of differential detector pairs, $\{I_{\text{Re}}^{+}, I_{\text{Re}}^{-}\}$ and $\{I_{\text{Im}}^{+}, I_{\text{Im}}^{-}\}$ (see Eq. (5) and **Methods**). **Figure 2B** displays representative wave processor outputs $\hat{F}_i$ overlaid with the target Berry–Popescu functions $F_i$ for several values of $a_i$. For low superoscillation strength (e.g., $a_i = 1.5$), the wave processor outputs closely track the ground truth across the entire input argument domain $[-t_{\text{roi}}, t_{\text{roi}}]$. As $a_i$ increases, the wave processor's approximation remains faithful near the origin, where the local superoscillation rate is highest, while deviations become more pronounced away from the center, where the target amplitude increases. To further quantify the reproduced superoscillatory behavior, **Fig. 2C** compares the local oscillation rate $k_{\text{local}}(t)$ calculated from the reconstructed optical outputs with those of the analytical BP functions for $a_i = 1.5, 2.5$, and 3.5. The reconstructed local oscillation rates closely follow the analytical values across the region of interest, reaching approximately $k_{\text{local}}(t) = a_i$ near the origin, well beyond the constituent harmonic bandlimit $k_{\max} = 1$. These results confirm that the diffractive wave processor reproduces not only the target functional profiles but also their characteristic locally superoscillatory behavior. To shed more light on these results, **Fig. 2D** contrasts the analytical structure of the target expansion with the spectral content of the diffractive output for superoscillation factors of $a_i = 1.5$, 2.5 and 3.5. The first column shows the exact expansion coefficients $A_{i,j}$ on the encoded harmonic grid $k_j \in [-1,1]$, giving the ground-truth weights of the bandlimited basis. Their vertical scales, spanning several orders of magnitude across the three rows, illustrate the rapid growth of the BP expansion coefficient magnitudes with increasing $a_i$, reflecting the strong cancellation required for superoscillations.

The intensity readout introduces an additional frequency scale that is useful for interpreting these results. As detailed above (Eq. (4)), square-law detection produces pairwise difference-frequency terms $k_p - k_q$. Since all encoded harmonics satisfy $k_p, k_q \in [-1,1]$, these quadratic cross-terms are globally restricted to

$$\left|k_p - k_q\right| \leq 2k_{\max} = 2. \tag{6}$$

The three examples in **Fig. 2D** therefore fall into two distinct regimes. For $a_i = 1.5$, the local oscillation rate exceeds the constituent and input-encoding bandlimit $|k| \leq 1$, but remains within the maximum difference-frequency range permitted by square-law detection. In contrast, for $a_i = 2.5$ and 3.5, the local oscillation rate exceeds not only the original constituent bandlimit but also the largest Fourier frequency that can arise from any individual quadratic cross-term, $\left|k_p - k_q\right| \leq 2$. These stronger superoscillatory cases of $a_i > 2$ in **Fig. 2D**, therefore, cannot be interpreted simply as the generation of a Fourier component at $k = \pm a_i$ through intensity mixing; rather, they require a locally superoscillatory response formed by interference among lower-frequency terms.

The remaining columns in **Fig. 2D** compare the absolute value of the DFT of the target functions' real and imaginary components, i.e., $\mathrm{Re}[F_i(t)], \mathrm{Im}[F_i(t)]$, with those of the corresponding differential outputs $\widehat{F}_{i,\mathrm{Re}} = I^+_{i,\mathrm{Re}} - I^-_{i,\mathrm{Re}}$ and $\widehat{F}_{i,\mathrm{Im}} = I^+_{i,\mathrm{Im}} - I^-_{i,\mathrm{Im}}$, all evaluated over $[-t_{\mathrm{roi}}, t_{\mathrm{roi}}]$. Because these spectra are evaluated over the restricted superoscillatory interval, they characterize the local oscillatory behavior rather than the global Fourier support of the underlying BP generating function or of the detected optical response. As $a_i$ increases from 1.5 to 3.5, the dominant spectral weight shifts progressively toward $k = \pm a_i$ (blue dashed lines), beyond the constituent bandlimit $|k| = 1$ of $G_i$ (red dashed lines). For $a_i = 2.5$ and 3.5, these local spectral features also lie beyond the quadratic difference-frequency bound $\left|k_p - k_q\right| \leq 2$. Their appearance in the finite-window spectrum should therefore be understood as a signature of the rapid local variation within the restricted interval. The last column reports the DFT magnitude of the residual error, $\widehat{F}_i - F_i$, which remains substantially below the signal spectrum at all three superoscillation strengths. Consistent with the increasing difficulty of achieving stronger superoscillations, the residual grows with $a_i$.

Having established the wave processing mechanism underlying superoscillatory function synthesis, we next quantify how the approximation fidelity scales with the two principal architectural resources of a diffractive wave processor: the number of trainable diffractive layers $L$, and the degrees of freedom (DOF) available per output channel, quantified by the ratio $r = \frac{N}{2N_i N_o} = \frac{N}{8N_{\text{base}}N_f}$, where $N_i = N_{\text{base}}$ represents the independent encoding pixels at the input plane, $N_o = 4N_f$ represents the independent pixels at the output plane (i.e., $\{I_{\text{Re}}^{+}, I_{\text{Re}}^{-}\}$ and $\{I_{\text{Im}}^{+}, I_{\text{Im}}^{-}\}$ for each superoscillatory function) and $N$ denotes the total number of optimizable phase features spread across all $L$ diffractive layers, with the diffractive features evenly distributed across the $L$ diffractive layers(*35*). **Figure 3A** reports the median MSE across the ensemble of $N_f = 100$ distinct complex-valued superoscillatory target functions as a function of $r$ for $L = 2, 4$ and $8$. Architectural depth emerges as the dominant resource: at a fixed feature ratio $r$, increasing $L$ from 2 to 8 reduces the median MSE by approximately an order of magnitude, a gain obtained purely by redistributing the same total trainable features ($N$) across more layers. Increasing $r$ at fixed depth, by contrast, yields comparatively modest gains that saturate beyond $r \approx 2$: for the deeper designs, the error falls steeply at small $r$ and thereafter shows little additional improvement up to the largest ratio examined, while for $L = 2$ it is essentially independent of $r$ across the entire range. This asymmetry indicates that deeper diffractive architectures achieve systematically lower errors. This behavior identifies diffractive depth as a resource for realizing demanding interference patterns within an unchanged native harmonic representation. Additional modulation–propagation stages improve approximation fidelity at matched total parameter count without introducing new argument-domain frequencies. The observed saturation with increasing lateral feature count therefore points to a practical allocation strategy: once additional lateral features provide diminishing gains, depth can be used to improve the accurate synthesis of strongly superoscillatory functions.

We further note that the shaded intervals in **Fig. 3A** remain broad at all depths, reflecting the heterogeneity of the target function ensemble rather than run-to-run optimization variance - a dependence resolved explicitly in **Fig. 3B**, which presents the per-function MSE as a function of the superoscillation parameter $a_i$ for each depth and $r$. Across all configurations, the error generally increases with $a_i$, confirming that stronger superoscillations are systematically harder to synthesize. Critically, the architectural configuration matters only in the demanding regime: for weakly superoscillatory targets ($a_i \lesssim 2$) all configurations generally achieve low approximation errors, whereas at $a_i \approx 5$ the configurations separate by more than an order of magnitude, from $\sim 1 \times 10^{-3} (L = 2)$ to $\sim 7 \times 10^{-5} (L = 8)$. For $L > 2$, larger $r$ values shift the curves downward, consistent with the trend of **Fig. 3A**. Beyond approximation fidelity, **Fig. 3C** further examines the reconstructed local oscillation rate at $t = 0$ for $L = 2, 4, 8$ and $r = 4$. The reconstructed local oscillation rates closely follow the analytical relation $k_{\text{local},i}(0) = a_i$ across the target ensemble, with increasing deviations at larger $a_i$, where stronger superoscillations demand more precise interference among the encoded harmonics. Stated differently, at larger $a_i$ values, the deeper architectures (e.g., $L = 8$) more faithfully reproduce the target local oscillation rates and this behavior identifies local-phase-rate fidelity as an additional measure of the architectural demands of strongly superoscillatory function synthesis. These results further confirm the ability of the diffractive processors to reproduce local oscillation rates beyond the constituent harmonic bandlimit, $k_{\max} = 1$. Representative best-, median- and worst-case approximations at $r = 2$ are shown in **Fig. 3D** for each depth. The examples selected as best cases correspond to $a_i \approx 1.2$–$2.1$ and those selected as worst cases to $a_i \approx 4.4$–$4.9$ at every depth, consistent with the increased approximation difficulty at larger $a_i$. For the selected examples, the MSE ratios between $L = 2$ and $L = 8$ are ~2× for the best cases and ~18× for the worst cases. Since these examples correspond to different target functions, these ratios represent illustrative comparisons rather than paired improvements for identical targets.

**Piecewise optical synthesis of superoscillations beyond the encoding basis**

The preceding BP results consider superoscillatory functions whose constituent harmonics remain within the input encoding band (i.e., $N_{\text{sup}} = N_{\text{base}}$), although their local oscillation rates can exceed that bandlimit; see e.g., $a_i > 2$ results in **Fig. 2C**. We next introduce spatial multiplexing of the functional representation, where different output channels are assigned to different portions of a target superoscillatory function, allowing spatial parallelism to resolve functional complexity. We demonstrate this architecture using targets whose harmonic content exceeds that of the native encoder, i.e., $\rho_h > 1$ or $N_{\text{sup}} > N_{\text{base}}$. In the architecture shown in **Fig. 4A** (with $L = 5$ optimized phase-only layers), the target domain is partitioned into $N_\Delta$ sub-intervals, each assigned to a dedicated output channel. Each sub-interval is expanded onto the full native encoding range, so that the wave processor learns a set of local responses that collectively represent the complete superoscillatory function. Spatial channel allocation thereby becomes a resource for superoscillatory functional complexity, reducing the portion of the target that each optical response must reproduce. For the $i$-th target function $F_i(t)$, the complete reconstruction is assembled over the target domain $[-t_{\text{roi}}, t_{\text{roi}}]$ by combining the interval-specific detector responses evaluated at their corresponding native encoding coordinates. Specifically, for a sub-interval of width $\Delta_{i,m}$ and center $t_{i,m}^{c}$, the target argument $t$ is mapped to the native encoding coordinate, $s = \frac{2t_{\text{roi}}}{\Delta_{i,m}}\left(t - t_{i,m}^{c}\right)$. The corresponding local target is given as: $F_{i,m}^{\text{local}}(s) = F_i\left(t_{i,m}^{c} + \frac{\Delta_{i,m}}{2t_{\text{roi}}}s\right)$. A target harmonic with wavenumber $\kappa_j$ therefore has wavenumber $\kappa_j \Delta_{i,m}/(2t_{\text{roi}})$ in the native encoding coordinate. Narrower intervals present slower functional variation to each optical channel, while the optical synthesis of channel-specific responses in parallel retains the structure of the complete target. For each query, the prescribed interval map determines the encoding coordinate and the detector used to obtain the corresponding response.

These intervals need not share a common width: each output channel is assigned its own local target, while the shared diffractive parameters are jointly optimized across channels, and non-uniform partitions can in principle allocate narrower tiles to regions where the target oscillates most aggressively. The results that we present below demonstrate the architecture using uniform tiling, which isolates the effect of the decomposition itself from that of the partition; the non-uniform case is examined separately later in **Fig. 6**. **Figure 4B** shows representative best, median, and worst-case approximation results for $N_f = 10$ distinct superoscillatory functions (simultaneously evaluated) with $N_{\text{base}} = 9$ and $N_{\text{sup}} = 36$, partitioned into $N_\Delta = 10$ intervals of equal width. Even the worst-case function achieves MSE on the order of $10^{-6}$. The per-interval MSE distribution (**Fig. 4C**) reveals uniform approximation quality across all $N_\Delta$ intervals, confirming that the shift-variant point spread function of the diffractive wave processor enables independent, high-fidelity approximation at each spatial tile without significant degradation across the output channels.

To further quantify the synthesized superoscillatory behavior, we calculate the absolute local oscillation rate using $|k_{\mathrm{local}}(t)| = |\mathrm{Im}\left[\frac{A'(t)}{A(t)}\right]|$, where $A(t)$ denotes the analytic signal obtained from the Hilbert transform of the real-valued function; see the **Methods** section for details. As shown in **Fig. 5**, all ten randomly generated target nonlinear functions exhibit genuine superoscillatory behavior, with $|k_{\mathrm{local}}(t)|$ exceeding the target constituent bandlimit $K_{\mathrm{tar}}^{(t)} = 2\pi N_{\mathrm{sup}} \approx 226$, reaching peak values of 520~849. The wave processor's approximations, computed via the Hilbert transform of the stitched output, faithfully reproduce the broad local oscillation-rate structure of the target functions, including the locations and relative magnitudes of superoscillatory peaks that far exceed $K_{\mathrm{tar}}^{(t)}$. Deviations between the target and approximated local oscillation rates are primarily confined to narrow zones around the zero amplitude regions, where $k_{\mathrm{local}}(t)$ diverges; these regions are inherently sensitive to small approximation errors. Notably, the native encoder employs only $N_{\mathrm{base}} = 9$ harmonics with a bandwidth of $K_{\mathrm{base}}^{(s)} = 2\pi R N_{\mathrm{base}} \approx 33.9$, yet the optically synthesized superoscillatory functions routinely achieve local oscillation rates exceeding $K_{\mathrm{tar}}^{(t)} \approx 226$, more than six times the encoder bandwidth. These results confirm the capability of the presented piecewise optical function synthesis architecture to approximate superoscillatory target functions with low MSE despite employing a native encoder bandwidth $K_{\mathrm{base}}^{(s)}$ that is much smaller than the target constituent bandlimit $K_{\mathrm{tar}}^{(t)}$.

Relaxing the equal-width constraint preserves the same behavior. In **Fig. 6A**, the domain is partitioned into intervals of unequal width, and the piecewise synthesis remains accurate across the ensemble, with the simultaneously synthesized superoscillatory reconstructions (through an optimized diffractive wave processor with $L = 5$) closely tracking the target across the sub-intervals despite their varying widths. These results reveal that wider tiles are systematically, though only modestly, harder to approximate, consistent with the increased effective target bandwidth with respect to the rescaled encoding coordinate, as wider intervals require each detector to reproduce a larger portion of the target function over the same encoding domain (**Fig. 6B**). The per-interval error map across all superoscillatory functions and tiles shows no systematic dependence on tile position, with the residual variation instead reflecting the local structure of each individual target function (shown in **Fig. 6C**). The absolute local oscillation rates of the target functions and the diffractive wave processor outputs are also reported in **Supplementary Fig. S1**, showing results similar to those obtained with uniform partitioning in **Fig. 5**. These results indicate that the piecewise decomposition is robust to the choice of partition, leaving room for adaptive, target-specific tiling strategies.

Beyond the geometry of the partition, the fidelity of the piecewise synthesis also depends on how the input variable is mapped into optical phase within each interval. This mapping is controlled by the encoding scaling factor ($R$), which globally scales the wavenumbers assigned to the optical encoder pixels according to $k_j = (j+1)2\pi R$. Accordingly, the phase encoded by the $j$-th pixel varies as $\phi_j(s) = k_j s$, and its total phase excursion is $\Delta\phi_j = (j+1)4\pi R t_{\text{roi}}$. Thus, at fixed $N_{\text{base}}$, $R$ does not change the number of encoded harmonic components; rather, it controls how rapidly the encoder phase varies with the native encoding coordinate $s$, and therefore the phase diversity available within each sub-domain. **Figure 4D** investigates the relationship between $R$ and the achievable MSE for varying numbers of target harmonics, $N_{\text{sup}}$; this analysis was conducted under the uniform-interval condition. For each $N_{\text{sup}}$, there exists an optimal range $R$ that balances two competing effects: too small of an $R$ limits the phase diversity available to the optical encoder. For larger $R$, the increased phase excursion can introduce ambiguity in the encoded inputs. At $N_{\text{sup}} = 16$, $R \approx 0.4$ achieves a mean MSE below $10^{-6}$. As $N_{\text{sup}}$ increases to 25, 36, 49 and 64, the optimal MSE degrades progressively, reflecting the growing difficulty of synthesizing higher-bandwidth content from a fixed $N_{\text{base}} = 9$ input encoding. The lower-error range shifts toward larger $R$ for higher-harmonic targets, consistent with the benefit of greater native phase variation.

In addition to the encoding scale $R$, the number of sub-intervals $N_\Delta$ provides another key degree of freedom in the piecewise architecture. Increasing $N_\Delta$ narrows the target domain assigned to each detector, thereby reducing the local synthesis difficulty, but this simultaneously increases the number of output channels/detectors that must share the same trainable feature budget. **Figure 4E** explores the mean MSE as a function of $N_\Delta$ at a fixed trainable feature budget $N = 5 \times 200 \times 200$ and $R = 0.6$. The error exhibits a minimum at $N_\Delta \approx 8 - 10$, rising by ~3-4 orders of magnitude on either side. The two branches originate from opposing constraints. For small $N_\Delta$, each detector is responsible for approximating the target function over a relatively wide sub-domain. The piecewise decomposition therefore provides only limited simplification of the original approximation task, and the system retains much of the difficulty associated with approximating the full superoscillatory target function set. For large $N_\Delta$, the increasing number of detector channels must share the same fixed trainable feature budget, limiting the available resources relative to the number of output channels; furthermore, the closer detector spacing additionally approaches the spatial resolution limit of the diffractive wave processor. The optimum therefore reflects a balance between the per-interval synthesis difficulty and the degrees of freedom available to the wave processor.

**Experimental validation**

We experimentally validate the spatially multiplexed superoscillation architecture using a visible-light phase-modulation set-up (**Fig. 7A,B**). A coherent laser at $\lambda = 520$ nm illuminates a phase-only SLM, and the modulated field is relayed through a 4f system to a CMOS image sensor. The experimental configuration uses $N_f = 1$ target function, $N_{\text{base}} = 4$ input harmonics, $N_{\text{sup}} = 9$ target harmonics, and $N_\Delta = 9$ sub-intervals with $R = 0.6$ and $t \in [-0.5, 0.5]$.

**Figure 7C** shows the optimized phase pattern displayed on the SLM $\phi_{\text{SLM}}$ when $s = 0.15$ (for visualization without loss of generality), comprising the phase summation of the input-dependent encoder $\phi_{\text{enc}}$ (dashed green box outlined) embedded within the input-independent decoder $\phi_d$. The encoder and decoder are co-located on the same SLM plane. The encoder provides fixed argument-dependent phase modulation, while the trainable decoder contributes an input-independent phase profile across the SLM, including the overlapping encoder pixels. Their phase contributions are summed to form the final SLM pattern. The illuminated decoder region outside the encoder also contributes an input-independent reference field, which coherently interferes with the encoded components at the detector. This introduces a zero-frequency component, extending the maximum intensity difference frequency to $2\pi R N_{\text{base}}$.

The experimental optimization was performed in two stages. First, the phase profile was numerically optimized using a differentiable wave-optical model based on the angular spectrum method (ASM), providing an initial design for experimental implementation. This optimized phase profile was then transferred to the physical SLM and further refined in situ using model-free proximal policy optimization (PPO)(*36*). Unlike simulation-based optimization, the experimental refinement relied directly on measured optical responses without requiring gradients to be computed through the physical system. The convergence of the experimental PPO refinement is presented in **Fig. 7D**. Starting from the simulation-optimized phase profile, the MSE decreases over ~580 epochs of in situ refinement, reaching a final value on the order of $10^{-3}$. Three representative epochs (marked on the curve) are highlighted, with their corresponding function reconstructions shown in **Fig. 7E**. At epoch 0 (early training), the optical output captures the coarse envelope of the target function. By epoch 300, the reconstruction tracks the target with substantially improved fidelity, and at epoch 580, the measured output closely follows the ground-truth superoscillatory function across all 9 intervals of the input domain $t \in [-0.5, 0.5]$, with the corresponding absolute local oscillation rate comparison shown in **Fig. 7F**. The alternating shaded bands in **Fig. 7E** delineate the $N_{\Delta} = 9$ sub-intervals, showing that the concatenated sensor readouts closely approximate the target across the full domain. Similar experimental results using $N_{\text{base}} = 4$ input harmonics to approximate a superoscillatory function with $N_{\text{sup}} = 6$ and $N_{\Delta} = 4$ are reported in **Supplementary Fig. S2**. These experiments demonstrate/validate the use of spatial optical channels to represent different portions of a superoscillatory function, translating the local-representation architecture into a programmable wave processor.

**Figure 7F** further compares the absolute local oscillation rates of the target function and the experimentally measured wave processor output at epoch 580. The target exhibits locally superoscillatory behavior, reaching a peak absolute local oscillation rate of ~226, well beyond its own constituent bandlimit of $K_{\text{tar}}^{(t)} = 2\pi N_{\text{sup}} = 18\pi$. The wave processor's measured output reproduces the broad local oscillation-rate structure, including regions beyond the target bandlimit, $K_{\text{tar}}^{(t)}$, although deviations occur near sharp peaks and regions of low analytic signal amplitude. Similar behavior is observed in the four-interval experiment (**Supplementary Fig. S2D**), where the measured local oscillation rate reaches ~75, nearly twice the corresponding target constituent bandlimit of $K_{\text{tar}}^{(t)} = 2\pi N_{\text{sup}} = 12\pi$. Importantly, these superoscillatory functions are synthesized using a native encoder with only $N_{\text{base}} = 4$ harmonics and a bandwidth of $K_{\text{base}}^{(s)} = 2\pi R N_{\text{base}} = 4.8\pi$ before the interval rescaling. Together, these results experimentally demonstrate the synthesis of superoscillatory behavior and the ability of the spatially multiplexed diffractive wave processor to synthesize nonlinear functions beyond the native encoding bandwidth.

## Discussion

This work connects superoscillatory wave interference with the architecture of optical computation. The BP results demonstrate the simultaneous approximation of a set of rapidly varying, complex-valued superoscillatory functions within the fixed spectral support of the optical readout, including targets whose local oscillation rates exceed those of every available detector harmonic. The spatially multiplexed architecture extends the superoscillatory functional representation by assigning locally rescaled portions of the target to different output channels. Together, these approaches connect the quantities that are often considered separately: the spectrum of the encoding, the local variation of the function, and the spatial resources of the wave processor.

In our analyses and results, propagation and phase modulation remain linear in the optical field, whereas the encoded field depends nonlinearly on the function argument. Intensity detection produces argument-domain difference frequencies, $k_p - k_q$, defining the global Fourier support of each detector response. Local variations beyond this support arise through superoscillatory interference over a restricted argument interval, rather than through the generation of a corresponding global Fourier component.

The DOF scaling analysis (**Fig. 3**) reveals two design principles for practical implementation. First, optical depth of the wave processor matters: increasing $L$ from 2 to 8 layers at fixed DOF ratio yields an order-of-magnitude reduction in MSE, indicating that the iterative application of mask modulation and free-space propagation provides qualitatively different wavefront shaping capabilities beyond mere parameter count. Second, the existence of a saturation point ($r \approx 2$) implies a practical design rule: in the tested architectures, increasing $r$ beyond ~2 offers diminishing returns, allowing the system designer to allocate resources toward architectural depth of the wave processor rather than lateral resolution.

The encoding scale and interval allocation jointly determine the local computational representation. Increasing $R$ increases the native encoder's phase variation, while reducing an interval's width reduces the target's variation with respect to the rescaled encoding coordinate. The finite optima in **Fig. 4D,E** show that these resources must be coordinated with the available diffractive features and output channels. Superoscillation functional complexity can therefore be addressed through the joint design of the encoding, the domain partition and the physical wave processor.

The visible-light experiments (e.g., **Fig. 7**) translate this representation into a compact optical implementation and confirm that the spatially multiplexed superoscillation architecture is physically realizable. Using an 8-bit phase-only SLM at visible wavelengths, the system successfully reconstructs a $N_{\mathrm{sup}} = 9$ harmonic target superoscillatory function from only $N_{\mathrm{base}} = 4$ input encoding harmonics, i.e., a target-to-encoding harmonic-count ratio of $\rho_h = N_{\mathrm{sup}}/N_{\mathrm{base}} = 2.25$. The convergence behavior during in situ optimization reveals that the coarse function envelope is captured early, while fine superoscillatory structure requires extended training;

as expected, the precise destructive interference required for superoscillation is harder to establish than gross amplitude modulation.

The presented architecture provides several directions for extending superoscillatory functional wave computation. Our results demonstrate multiple functions and prescribed nonuniform partitions. Adaptive partition selection could allocate narrower intervals and additional optical resources to regions with more demanding local structure. Extension to multidimensional arguments would allow detector channels to represent local regions of a higher-dimensional function domain, while wavelength-dependent channel allocation could provide an additional multiplexing dimension. Scaling these representations will require coordinated design of the encoder area, diffractive features and detector layout. These directions point toward optical computing systems whose physical resources are organized around the local structure of the functions they evaluate.

The broader significance of these results lies in a distinct route to beyond-bandwidth function approximation. Rather than enlarging the encoded Fourier-feature set through nonlinear frequency mixing at the detector, the present framework exploits the local nature of superoscillations: a restricted region of a target function can exhibit variations faster than the highest frequency available in the input encoding while remaining accurately approximated through coherent interference among the encoded components. By further partitioning the target domain into multiple sub-intervals, the spatially multiplexed architecture extends this principle to functions whose global spectral content exceeds the input-encoding bandwidth, with each detector responsible for a locally tractable approximation task. Intensity detection provides the optical readout, but the key mechanism enabling the beyond-bandwidth approximation is the restriction of the approximation domain rather than the generation of additional Fourier components through square-law mixing. This approach therefore enables superoscillatory function synthesis without explicit calculation of the ill-conditioned expansion coefficients associated with conventional superoscillation design or high-intensity nonlinear optical interactions.

This work addresses a fundamentally different problem from conventional optical superoscillation studies. Rather than designing an optical element to generate a predetermined superoscillatory spatial field or intensity pattern, we implement a superoscillatory function as a physical input–output mapping: a continuously variable argument is encoded optically, and the same fixed diffractive wave processor produces an optical response that approximates the target function after normalization, including the real and imaginary components of complex-valued superoscillatory functions. Crucially, the frequencies that characterize these functions are frequencies with respect to the abstract input coordinate and should not be confused with the transverse spatial frequencies of the propagating optical field. The optical waves inside the processor remain fully governed by diffraction and contain only physically allowed propagating modes; the rapid variations arise instead in the dependence of the measured output on the queried function argument. This separation between physical spatial bandwidth and functional bandwidth allows a diffraction-limited optical system to realize input–output responses that locally vary faster than the frequencies represented in the input encoding, and, through spatial multiplexing, to evaluate targets whose harmonic content extends beyond that encoding bandwidth. The advance is therefore not the creation of another static superoscillatory optical waveform, but the physical realization of superoscillation as a dynamically queried computational function over a continuous input domain, which constitutes a distinct use of superoscillation that extends optical function evaluation beyond the bandwidth of its encoded representation.

In conclusion, we realize superoscillation as a computational input–output relation of a wave processor and demonstrate a spatially multiplexed architecture for extending optical function evaluation beyond an individual encoder's native harmonic content. Learned coherent interference enables approximation of rapid local variations, including the real and imaginary components of complex-valued superoscillatory functions. Local coordinate expansion and spatial channel allocation provide a distributed representation of the complete target, validated through visible-light experiments. These results establish functional superoscillation as a principle for designing analog optical processors and connect the spectral structure of mathematical functions to the physical organization of wave-based computation.

## Methods

### Physical forward model and wave processor architecture

The wave processor consists of $L$ trainable phase-only diffractive layers separated by free-space propagation regions of distance $dz$. An initial free-space propagation precedes the first diffractive layer, yielding a total of $L+1$ propagation segments and $L$ trainable layers. For the numerical studies presented in **Figs. 2–6**, simulations are performed at a pixel pitch of $dx = dy = 300$ nm and an operating wavelength of $\lambda = 550$ nm. The inter-layer propagation distance is fixed at $dz = 10\lambda = 5.5\ \mu$m unless otherwise mentioned.

Free-space propagation between successive diffractive layers is modeled using the angular spectrum method (ASM). For a monochromatic optical field propagating a distance $dz$, the free-space transfer function at spatial frequency $(f_x, f_y)$ is defined as,

$$H(f_x, f_y) = \begin{cases} \exp\left(\mathrm{i}\frac{2\pi}{\lambda} dz \sqrt{1-(\lambda f_x)^2 - (\lambda f_y)^2}\right), & f_x^2 + f_y^2 < \frac{1}{\lambda^2} \\ 0, & f_x^2 + f_y^2 \geq \frac{1}{\lambda^2} \end{cases}, \tag{7}$$

where $\lambda$ is the illumination wavelength. For propagating waves within the non-evanescent spatial-frequency band $\left(f_x^2 + f_y^2 < \frac{1}{\lambda^2}\right)$, the transfer function satisfies $|H(f_x, f_y)| = 1$. The output field after propagation is computed as $E_{\text{out}} = \mathcal{F}^{-1}\{H \cdot \mathcal{F}\{E_{\text{in}}\}\}$ using two-dimensional fast Fourier transforms.

Each diffractive layer maintains a trainable phase-only modulation profile of dimension $N_{\text{mask}} \times N_{\text{mask}}$ pixels (pixel size 300 nm). The physical phase $\varphi$ is parameterized through a latent real-valued variable $\theta$ via:

$$\varphi = (\sin\theta + 1) \cdot \pi \tag{8}$$

which maps any $\theta \in \mathbb{R}$ to $\varphi \in [0,2\pi]$ with bounded gradients. The resulting mask transmission function is $M = e^{i\varphi}$, satisfying $|M| = 1$ and ensuring physical realizability with phase-only SLMs or fabricated diffractive elements.

**Superoscillatory target function generation**

For the Berry–Popescu superoscillatory functions, the target ensemble consists of $N_f$ functions (variable with respect to $t$), each parameterized by a distinct superoscillation strength $a_i$:

$$G_i(a_i, n; t) = \left(\cos\frac{t}{n} + ia_i \sin\frac{t}{n}\right)^n = \sum_{j=0}^{n} A_{i,j}\,(a_i)\, e^{ik_j t} \tag{9}$$

where $i = 1, \dots, N_f$ indexes the target function and $j = 0,1, \dots, n$ indexes the phase encoding pixel on the input SLM. The wavenumbers assigned to each pixel are:

$$k_j = 1 - \frac{2j}{n}, j = 0,1, \dots, n \tag{10}$$

with $n = N_{\text{sup}} - 1$, yielding $N_{\text{sup}}$ frequencies uniformly spaced across $[-1, +1]$. The corresponding Fourier coefficients are given by:

$$A_{i,j} = \binom{n}{j}\left(\frac{1 + a_i}{2}\right)^{n-j}\left(\frac{1 - a_i}{2}\right)^{j} \tag{11}$$

The super-oscillation parameter $a_i$ is sampled uniformly from $[1,5]$. The superoscillatory functions $F_i(t)$ occur within the central region of interest bounded by $[-t_{\text{roi}}, t_{\text{roi}}]$, given by:

$$F_i(t) = G_i(a_i, n; t)[u(t + t_{\text{roi}}) - u(t - t_{\text{roi}})] \tag{12}$$

where $u(\cdot)$ is the unit step function. Despite all constituent frequencies $k_j$ lying within $[-1, +1]$, the BP function has a local oscillation rate of $k_{\text{local}}(t) = \frac{a_i}{\cos^2(t/n)+a_i^2 \sin^2(t/n)}$, yielding $k_{\text{local}}(0) = a_i$. Thus, the function locally oscillates faster than its highest constituent harmonic by a superoscillation factor of $a_i$. We are only concerned with the approximation within a region of interest $[-t_{\text{roi}}, t_{\text{roi}}]$, where $t_{\text{roi}} = 2$ for BP superoscillatory functions unless otherwise mentioned. The ground-truth values are computed in float64 precision because at high $a_i$ the coefficients $A_{i,j}$ span many orders of magnitude. In the results reported in **Fig. 2** and **Fig. 3**, $N_{\text{base}} = 16$ and $N_f = 100$ unless otherwise mentioned.

For the piecewise beyond-bandwidth superoscillation synthesis, $N_f$ independent target functions are generated as:

$$F_i(t) = \text{Re}\left[\sum_{j=0}^{N_{\text{sup}}-1} w_{i,j}\ \exp(i\kappa_j t)\right][u(t+t_{\text{roi}}) - u(t-t_{\text{roi}})] \tag{13}$$

where $i = 1, \dots, N_f$, $\kappa_j = 2\pi(j+1)$ denotes the target harmonic wavenumbers, distinct from the input encoding wavenumbers $k_j$ defined in Eq. (15), and $w_{i,j}$ are complex-valued random weights drawn from a fixed random seed to ensure reproducibility. The number of target harmonics $N_{\text{sup}}$ deliberately exceeds the number of input encoding harmonics $N_{\text{base}}$ (i.e., $\rho_h = N_{\text{sup}}/N_{\text{base}} > 1$), requiring the system to synthesize spectral content beyond its native input bandwidth. Each target function is approximated within a region of interest $[-t_{\text{roi}}, t_{\text{roi}}]$, where $t_{\text{roi}} = 0.5$ unless otherwise mentioned. In the results reported in **Fig. 4**, $N_f = 10$ and $N_{\text{base}} = 9$ unless otherwise mentioned. The target function domain $[-t_{\text{roi}}, t_{\text{roi}}]$ was then partitioned into $N_\Delta$ non-overlapping function-dependent sub-intervals with width $\Delta_{i,m}$, where $\sum_{m=1}^{N_\Delta} \Delta_{i,m} = 2t_{\text{roi}}$.

## Supplementary Information includes:

- **Supplementary Figs. S1-S2**
- **Methods on:**
  - **Input phase encoding of superoscillatory function input arguments**
  - **Optical readout of the superoscillatory function approximation**
  - **Loss function and training procedure**
  - **Experimental set-up**

## Figures

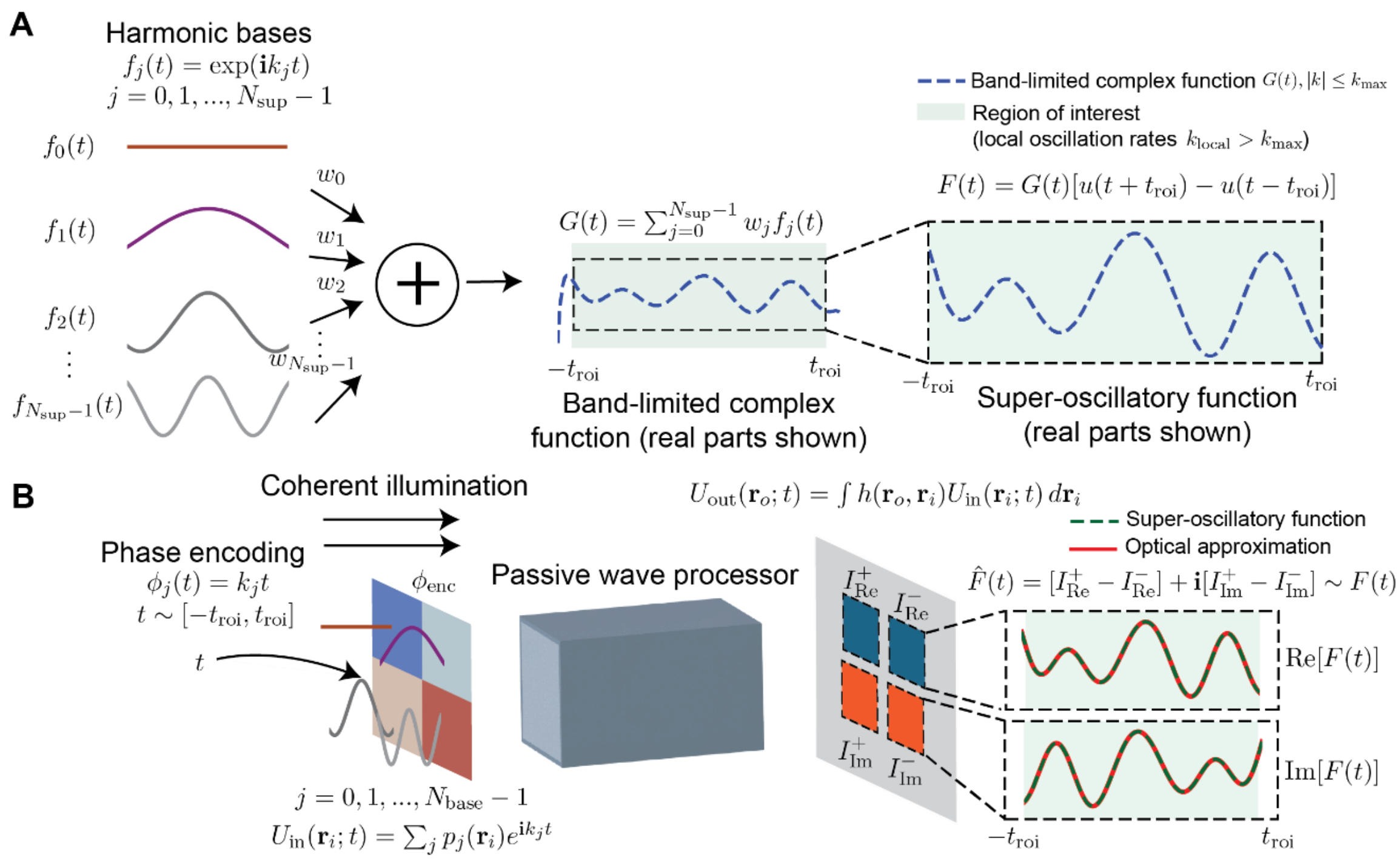

**Fig. 1. Superoscillatory function approximation using a passive wave processor.** **A** Synthesis of superoscillatory functions. A set of $N_{\mathrm{sup}}$ harmonic basis functions, $f_j(t) = \exp(\mathrm{i}k_j t)$ with $j = 0,1,\ldots,N_{\mathrm{sup}} - 1$, is combined with weights $w_j$ to form the band-limited function $G(t) = \sum_{j=0}^{N_{\mathrm{sup}}-1} w_j f_j(t)$. The weights are selected such that, within the region of interest $[-t_{\mathrm{roi}}, t_{\mathrm{roi}}]$ (shaded), the local oscillation frequency of $G(t)$ reaches $k_{\mathrm{local}} > k_{\max}$, i.e., the function locally oscillates faster than the bandlimit set by its own highest constituent harmonic. The target superoscillatory function is obtained by restricting $G(t)$ to this region, i.e., $F(t) = G(t)[u(t + t_{\mathrm{roi}}) - u(t - t_{\mathrm{roi}})]$. **B** Optical approximation of the target superoscillatory function. Under coherent illumination, the scalar input $t$ is mapped onto the encoder phase profile $\phi_{\mathrm{enc}}(t)$, in which each of the $\sqrt{N_{\mathrm{base}}} \times \sqrt{N_{\mathrm{base}}}$ input pixels carries a phase $\phi_j(t) = k_j t$ proportional to its assigned wavenumber, so that the input plane encodes the input harmonic basis functions. The resulting wavefront propagates through a passive diffractive wave processor, which synthesizes the output field $U_{\mathrm{out}}(\boldsymbol{r}_o; t)$. The real (Re) and imaginary (Im) components of the complex-valued superoscillatory target function are recovered using two sets of differential intensity detectors. Dashed curves denote the ground-truth superoscillatory function (Re and Im parts) and solid curves denote the optical approximation.

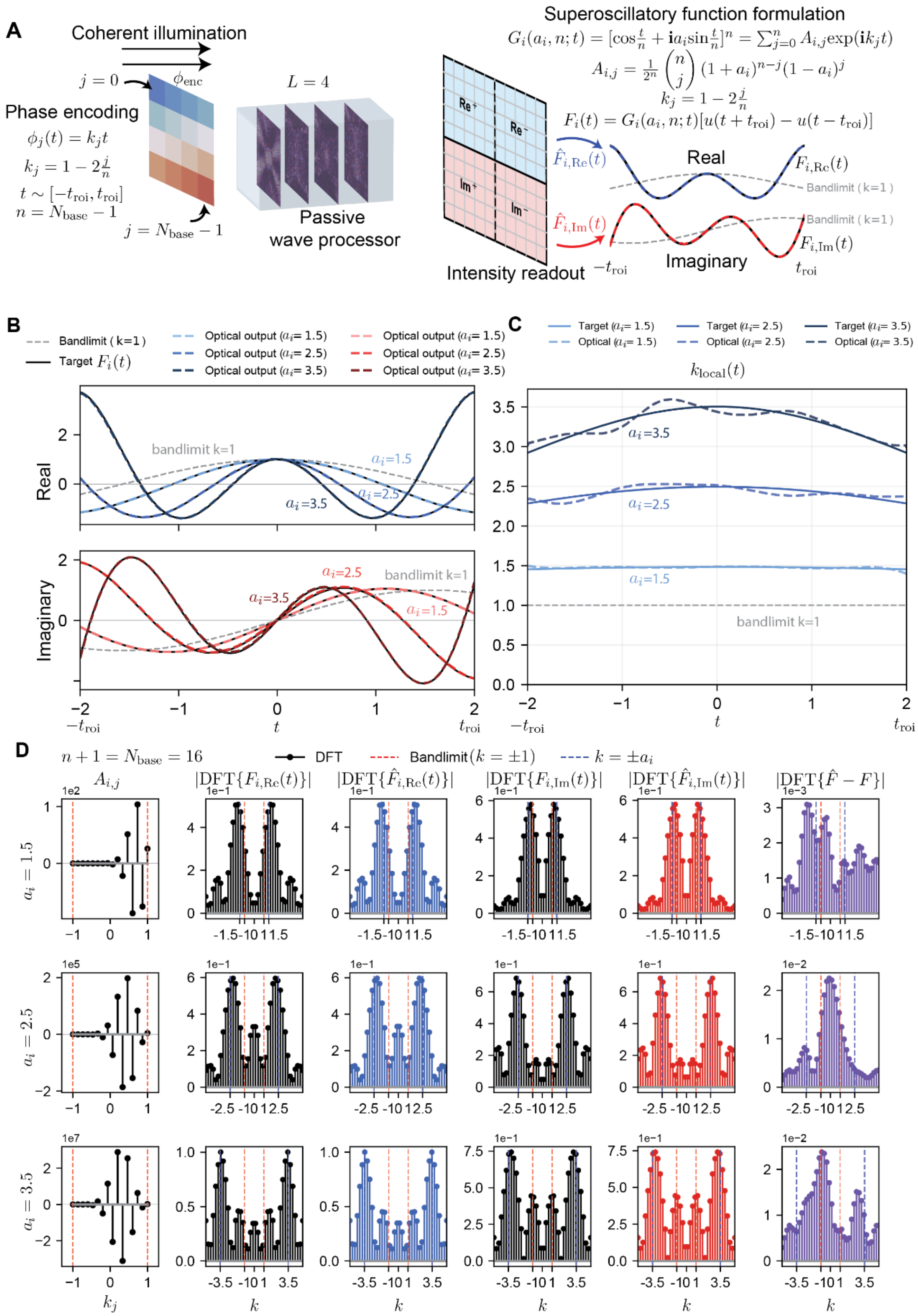

**Fig. 2. Approximation of Berry–Popescu (BP) superoscillatory functions using a diffractive wave processor.** **A** Schematic of the architecture and target function formulation. The scalar input

$t$ is encoded as the phase profile $\phi_{\text{enc}}$ with wavenumbers $k_j = 1 - 2j/n$ spanning $[-1,1]$, where $n = N_{\text{base}} - 1$. The Berry–Popescu generating function $G_i(a_i, n, t)$ is band-limited to $|\ k\ | \leq 1$ yet oscillates locally at frequency $a_i > 1$ near $t = 0$; the target is $F_i(t) = G_i(a_i, n, t)[u(t + t_{\text{roi}}) - u(t - t_{\text{roi}})]$. The output plane is partitioned into four detector regions ($\text{Re}^+$, $\text{Re}^-$, $\text{Im}^+$, $\text{Im}^-$), recovering the signed real and imaginary parts of the superoscillatory function, $\hat{F}_{i,\text{Re}}(t)$ and $\hat{F}_{i,\text{Im}}(t)$, from intensity-only measurements. **B** The outputs of a diffractive wave processor with $L = 4$ diffractive layers (coloured dashed curves) overlaid with the targets (black solid curves) for $a_i = 1.5, 2.5$ and 3.5. Grey dashed curves indicate the highest frequency accessible within the bandlimit ($k = 1$). **C** Local oscillation rate $k_{\text{local}}(t)$ for $a_i = 1.5, 2.5$, and 3.5, calculated from the target functions analytically using $k_{\text{local}}(t) = \text{Im}[G'(t)/G(t)]$ (solid curves) and the corresponding reconstructed optical outputs (dashed curves). The grey dashed line marks the constituent harmonic bandlimit $k_{\max} = 1$, highlighting the locally superoscillatory behavior beyond the encoded harmonic range. **D** Spectral analysis at $n + 1 = N_{\text{base}} = 16$. The first column shows the exact expansion coefficients $A_{i,j}$, illustrating the rapid growth of the expansion coefficient magnitudes with increasing $a_i$. The next four columns compare the DFT magnitudes of the target real and imaginary parts with those of the diffractive processor outputs, evaluated over $[-t_{\text{roi}}, t_{\text{roi}}]$; the last column shows the DFT magnitude of the residual $\hat{F} - F$. Red dashed lines mark the bandlimit $k = \pm 1$ and blue dashed lines mark the local super-oscillation frequency $k = \pm a_i$ . Results use $N_f = 100$ target functions with $a_i$ sampled uniformly from $[1, 5]$.

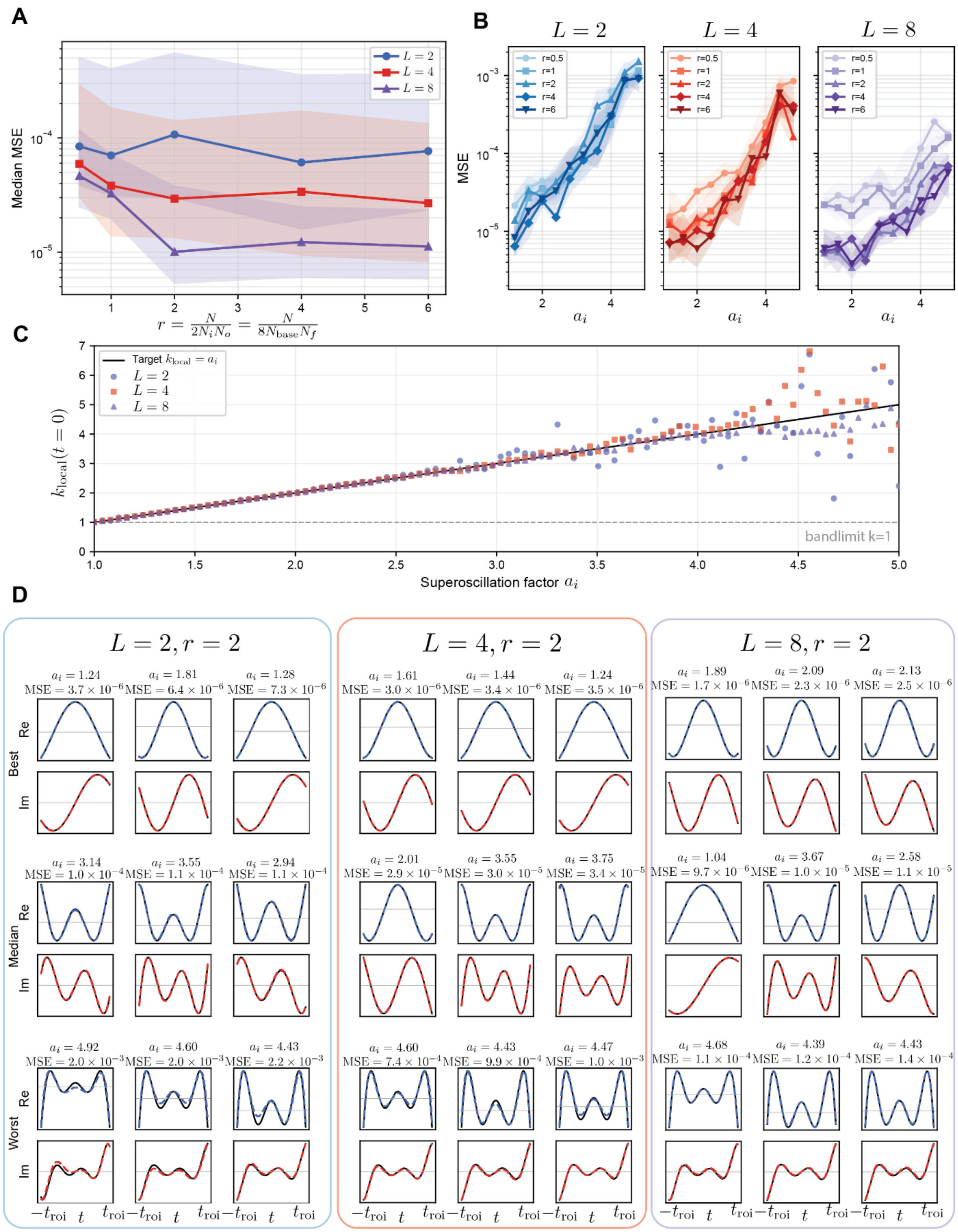

**Fig. 3. Scaling of the superoscillatory function approximation fidelity as a function of the trainable feature budget and the number of diffractive layers. A** Median MSE across the

ensemble of $N_f = 100$ Berry–Popescu target functions as a function of $r = N/(2N_i N_o) = N/(8N_{\text{base}} N_f)$, for $L = 2$, 4 and 8 diffractive layers. $N = L \times N_{\text{mask}}^2$ is the total number of trainable phase features summed over all layers. Shaded bands indicate the spread across the target function ensemble ($N_f = 100$). **B** Per-function MSE as a function of the super-oscillation parameter $a_i$, shown for each depth and for $r = 0.5$, 1, 2, 4 and 6. Shaded bands indicate the spread across independently trained models. **C** Reconstructed local oscillation rate $k_{\text{local}}(t = 0)$ resulting from the diffractive wave processor output as a function of the superoscillation parameter $a_i$ for $L = 2, 4$ and 8 diffractive layers with $r = 4$. The black solid line denotes the target local oscillation rate $k_{\text{local}}(t = 0) = a_i$, while the colored markers represent the corresponding optical approximations. The grey dashed line indicates the constituent harmonic bandlimit $k_{\max} = 1$. **D** Representative best-, median- and worst-case approximations at $r = 2$ for each depth ($L$), showing the real (blue) and imaginary (red) parts of the diffractive output as dashed curves overlaid with the target functions (black solid curves). The corresponding $a_i$ and MSE values are indicated above each panel.

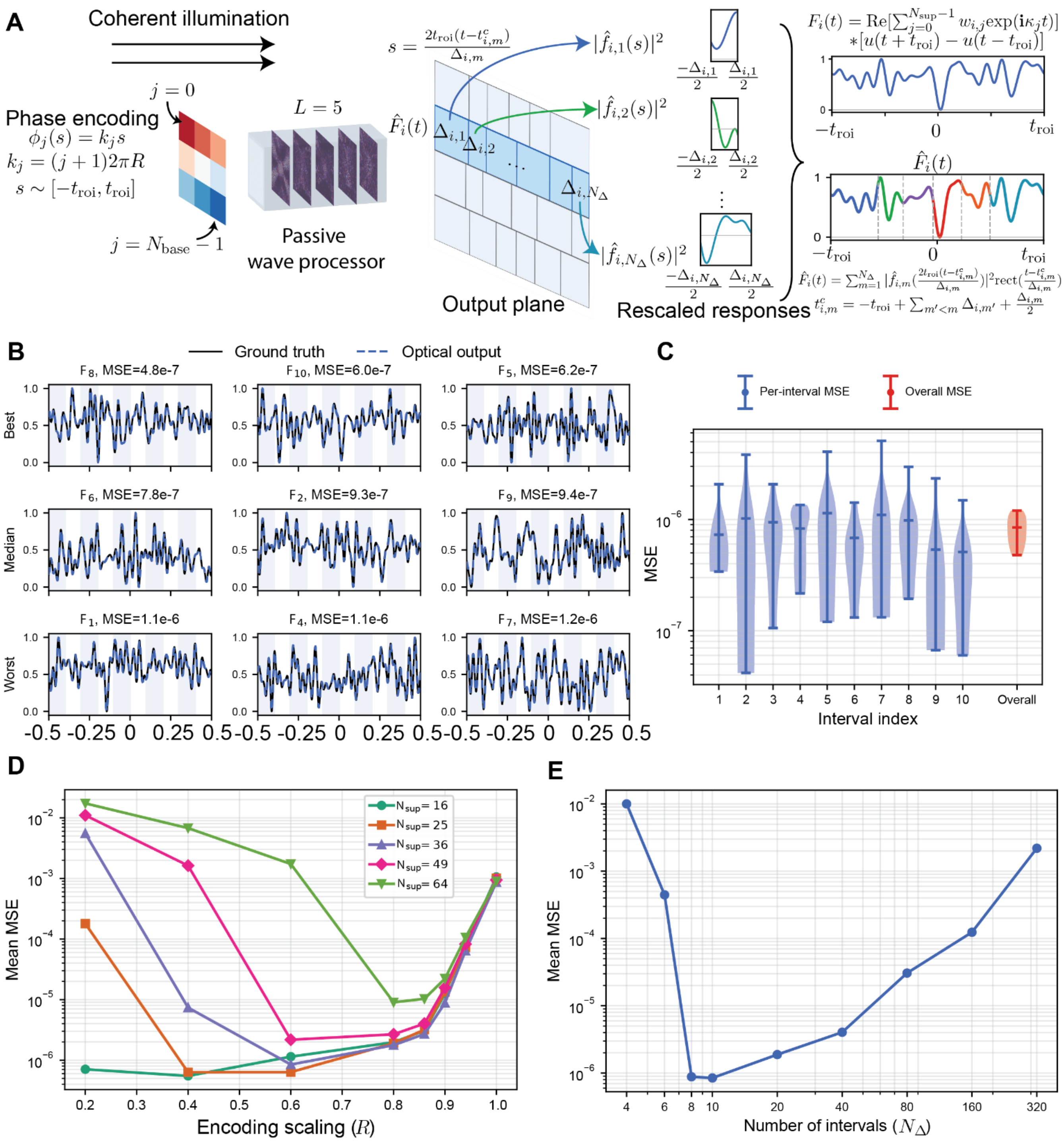

**Fig. 4. Piecewise optical synthesis of beyond-bandwidth superoscillatory functions. A** Schematic of the architecture. The native encoding coordinate $s$ is encoded as the phase profile $\phi_j(s) = k_j s$ with $k_j = (j+1)2\pi R$, $s \in [-t_{\mathrm{roi}}, t_{\mathrm{roi}}]$ where $R$ denotes the encoding scaling factor. The target domain $[-t_{\mathrm{roi}}, t_{\mathrm{roi}}]$ is partitioned into $N_\Delta$ sub-intervals, each assigned to a dedicated output detector, and the concatenated readouts reconstruct the full function as $\hat{F}_i(t) =$

$\sum_{m=1}^{N_\Delta} |\hat{f}_{i,m}\left(\frac{2t_{\mathrm{roi}}(t-t_{i,m}^{c})}{\Delta_{i,m}}\right)|^2 \ \mathrm{rect}\left(\frac{t-t_{i,m}^{c}}{\Delta_{i,m}}\right)$, where $t_{i,m}^{c}$ is the centre of interval $m$. The target harmonic count $N_{\mathrm{sup}}$ exceeds the number of encoded harmonics $N_{\mathrm{base}}$, so that the synthesis extends beyond the native input bandwidth. **B** Best, median and worst approximation results of an optimized diffractive wave processor ($L = 5$) for $N_f = 10$ target superoscillatory functions using $N_\Delta = 10$ with $N_{\mathrm{base}} = 9$, $N_{\mathrm{sup}} = 36$ and $R = 0.6$. Black solid curves denote the ground truth and blue dashed curves the optical output; alternating shaded bands delineate the $N_\Delta$ sub-intervals. **C** Per-interval MSE distribution across the target ensemble, shown alongside the overall MSE. **D** Approximation MSE as a function of the encoding scaling $R$ for target harmonic counts $N_{\mathrm{sup}} = 16$, 25, 36, 49 and 64, at fixed $N_{\mathrm{base}} = 9$, which correspond to target-to-encoding harmonic ratios of $\rho_h \approx 1.78, 2.78, 4.00, 5.44$, and 7.11, respectively. **E** Approximation MSE as a function of the number of intervals $N_\Delta$ at a fixed trainable feature budget $N$.

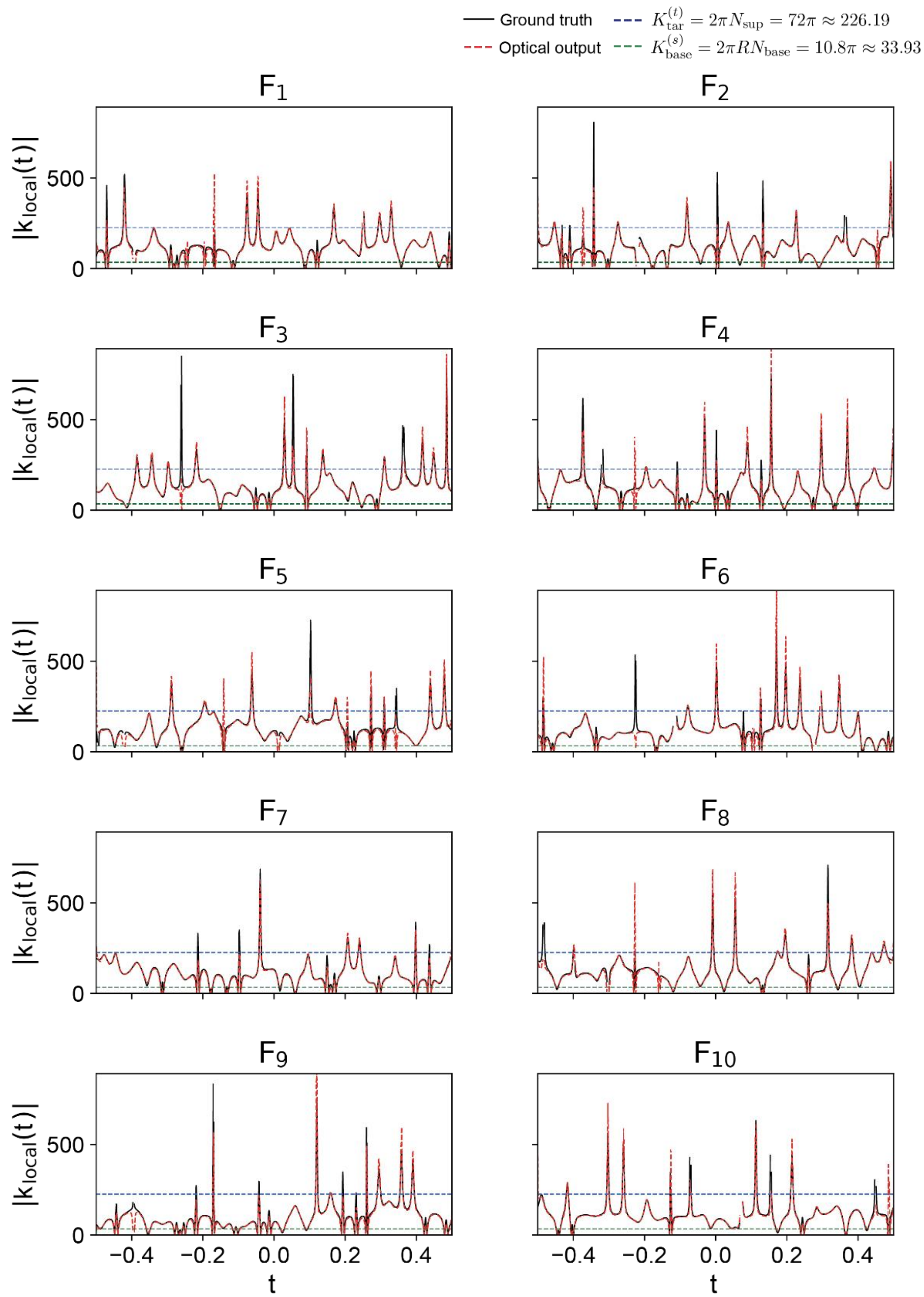

**Fig. 5. Absolute local oscillation rates $|k_{\mathrm{local}}(t)|$ of the target superoscillatory functions and the diffractive wave processor results.** Black solid and red dashed lines indicate the absolute local oscillation rates of the target functions and the diffractive wave processor results shown in **Fig. 4B**, respectively. The blue dashed line marks the target constituent bandlimit $K_{\mathrm{tar}}^{(t)} = 2\pi N_{\mathrm{sup}} = 72\pi$, while the green dashed line indicates the native encoding bandwidth of the wave

processor, i.e., $K_{\mathrm{base}}^{(s)} = 2\pi R N_{\mathrm{base}} = 10.8\pi$, before the interval rescaling. Local oscillation rates are evaluated with respect to the target argument $t$. Also see **Supplementary Fig. S1** for the non-uniform domain partitioning results.

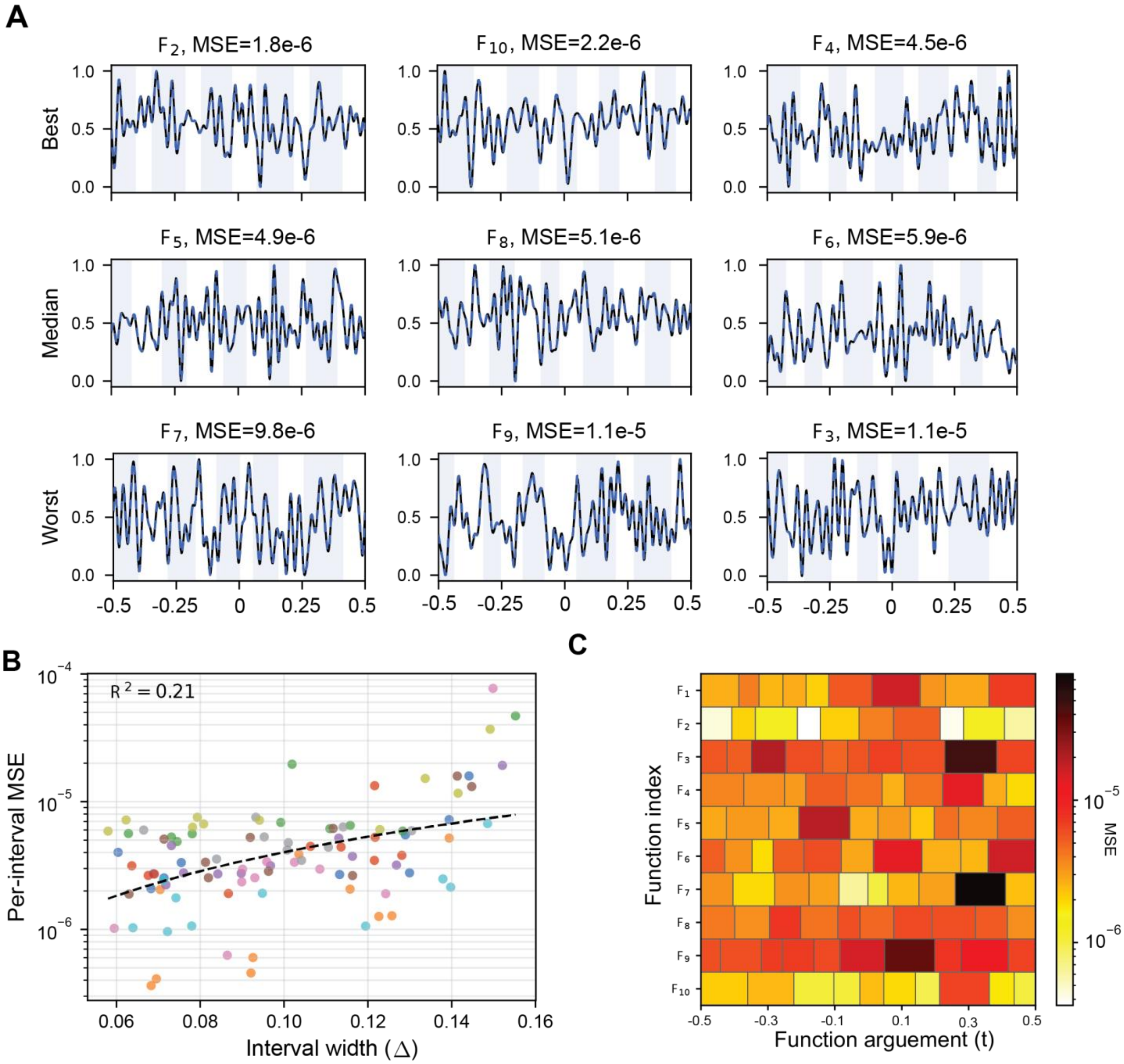

**Fig. 6. Piecewise optical synthesis of beyond-bandwidth superoscillatory functions with non-uniform domain partitioning. A** Representative best, median and worst approximation results of an optimized diffractive wave processor ($L = 5$) among $N_f = 10$ target functions partitioned into $N_\Delta = 10$ sub-intervals of unequal width with $N_{\text{base}} = 9$, $N_{\text{sup}} = 36$, $\rho_h = 4$ and $R = 0.6$. Black solid curves denote the ground truth and blue dashed curves denote the optical approximation; alternating shaded bands delineate the sub-intervals. **B** Per-interval MSE versus sub-interval width, pooled across all functions and sub-intervals, with each color denoting one target function; the dashed line is a least-squares fit ($R^2 = 0.21$). **C** Per-interval MSE across all target functions under nonuniform domain partitioning. Each cell represents a sub-interval, with its horizontal extent proportional to the interval width and its color indicating the corresponding MSE.

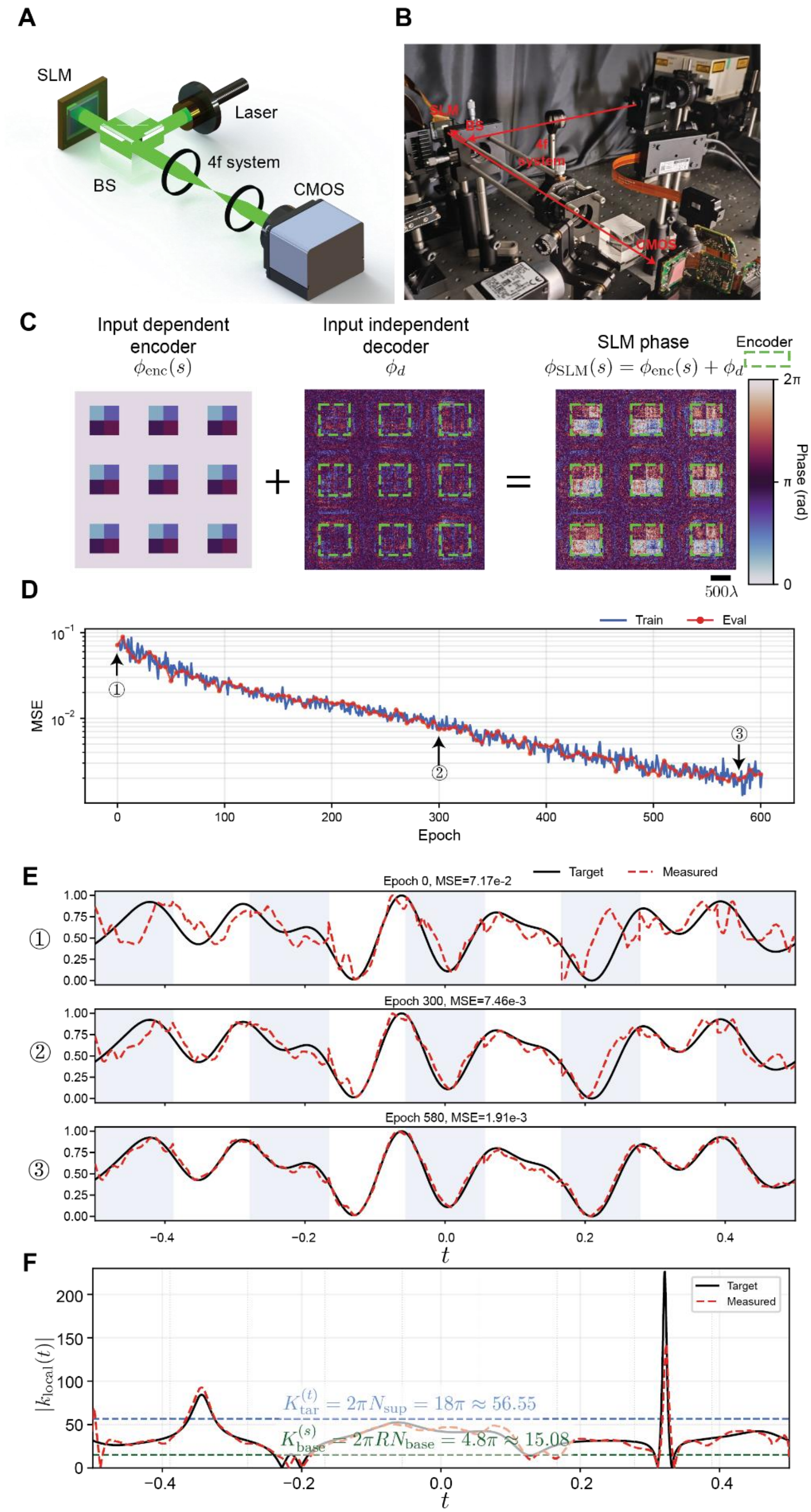

**Fig. 7. Experimental demonstration of the optical synthesis of beyond-bandwidth superoscillatory functions. A,B** Schematic and photo of the optical set-up. A laser beam at $\lambda =$

520 nm is redirected by a beam splitter (BS) and modulated by a phase-only SLM; the reflected field propagates through a 4f system acting as a relay and is captured by a CMOS camera. **C** Phase distributions implemented on the SLM for the nine processor units arranged in a 3 × 3 mosaic, shown at $s = 0.15$ for visualization. From left to right, the panels show the argument-dependent encoder phase $\phi_{\mathrm{enc}}\ (s)$, the input-independent decoder phase $\phi_d$ and the resulting SLM phase $\phi_{\mathrm{SLM}}(s) = \phi_d + \phi_{\mathrm{enc}}(s)$. The encoder and decoder are co-located on the same SLM plane: within the encoder regions (green dashed boxes), the same physical SLM pixels simultaneously carry the argument-dependent encoding and the decoder phase modulation, whereas outside these regions only the decoder phase is applied. **D** Training and evaluation MSE during model-free in situ optimization over 600 epochs. **E** Measured optical outputs (red dashed curves) overlaid with the target superoscillatory function (black solid curve) at epochs 0, 300 and 580, showing the progressive improvement enabled by in situ training. Alternating shaded bands delineate the $N_\Delta$ sub-intervals, each read out by a dedicated detector region on the CMOS sensor. **F** Absolute local oscillation rate $|k_{\mathrm{local}}(t)|$ of the target function (black solid curve) and experimentally measured optical output at epoch 580 (red dashed curve), calculated from their respective analytic signals. The blue dashed line marks the target constituent bandlimit $K_{\mathrm{tar}}^{(t)} = 2\pi N_{\mathrm{sup}} = 18\pi$, while the green dashed line indicates the native encoding bandwidth of the wave processor, i.e., $K_{\mathrm{base}}^{(s)} = 2\pi R N_{\mathrm{base}} = 4.8\pi$, before the interval rescaling. Local oscillation rates are evaluated with respect to the target argument $t$. The experimental configuration uses $N_f = 1$ target function, $t_{\mathrm{roi}} = 0.5$, $N_{\mathrm{base}} = 4$ input harmonics, $N_{\mathrm{sup}} = 9$ target harmonics (i.e., a target-to-encoding harmonic ratio of $\rho_h = 2.25$) and $R = 0.6$ with $N_\Delta = 9$; see **Methods** for details.